\def\year{2018}\relax
\documentclass[letterpaper]{article} 
\usepackage{aaai18}  
\usepackage{times}  
\usepackage{helvet}  
\usepackage{courier}  
\usepackage{url}  
\usepackage{graphicx}  
\begin{document}
%
\title{Related Fact Checks: a tool for combating fake news}
\author{Sreya Guha\\
Castilleja High School\\
Palo Alto, California\\
}
\maketitle

\begin{abstract}
The emergence of "Fake News"
 and misinformation via online news and social media has spurred  
an interest in computational tools to combat this phenomenon. In this paper we present a new "Related Fact Checks" service, which can help a 
reader critically evaluate an article and make a judgment on its veracity by bringing up fact checks that are relevant to the article. 
We describe the core technical problems that need to be solved in building 
 a "Related Fact Checks" service, 
and present results from an evaluation of an implementation. 

\end{abstract}



\section{Introduction}

In this paper, we explore the task of ameliorating the problem of `fake news' within an Information Retrieval framework, building upon prior research in Computer Science. Specifically, we introduce  a `Related Fact Checks' service to combat fake news. We discuss technical challenges in the implementation of this service, describe the browser extension through which the service is accessed, and present precision and recall numbers from an evaluation of the results provided by the service.

Recent years have witnessed the rising influence of fake news. Fake news articles, defined by the New York Times as `made-up [stories] with an intention to deceive' \cite{nytfakenews}, are one of the most serious challenges facing the news industry today. Rising to prominence in the 2016 United States Presidential Election, fake news has affected over 50\% of the voting population \cite{allcott2017social}. Fake news is having a broad impact, beyond politics. For example, according to Hogan \cite{IrishVaccine}, fake news about the effects of vaccines are causing a drop in the vaccination rate amongst certain groups.

Though the prevalence of fake news is an ethics and journalism issue, the impact of fake news is in large part because of artifacts such as Social Networks that have their roots in Computer Science. Consequently, combating fake news has become a very active area of research in Computer Science over the last year. While automatically fact checking articles remains the ultimate goal of many Computer Scientists, it is now recognized that this is a very complex task that requires not just extremely sophisticated text understanding, but also a level of common sense and world knowledge that computational tools currently do not have. This is further complicated by the fact that the veracity of articles cannot be defined in a binary sense of fake or not fake. Stories are often more complex and do not completely fit a simple binary classification. For example, in a 2017 interview with the Wall Street Journal \cite{wsjtrump}, President Trump claimed that Korea used to be a part of China. Numerous sites such as dailybanter.com and QZ.com declared his claim to be false. However, according to the Washington Post \cite{wpkorea}, President Trump's claim, while not capturing the entire story, was not completely false. Indeed, many fact checkers have a scale of accuracy rather than a simple binary distinction. PolitiFact \cite{politifact} for example, assigns each claim one of 6 rulings that range from `True' to `Pants on Fire'. The Washington Post \cite{washingtonpost} assigns each claim one of 5 rulings, ranging from `Geppetto Checkmark' (True) to `4 Pinnochios' (False). The subtlety captured in a fact checking article cannot be replicated by today's programs.

Due to the rise of fake news, fact checking is becoming more prevalent. According to the Duke Reporter's Lab \cite{dukereporterslab}, there are over 100 active fact checking websites in the world.  In 2015, Schema.org \cite{schema}, working with the Duke Reporters Lab, released vocabulary, including a new type, $ClaimReview$ for adding Semantic markup to fact checks, to make them more portable and accessible. As of September 2017, according to our crawl, there are over 5,000 fact checks with ClaimReview markup. 

\subsection{Related Fact Checks Service}
 The news ecosystem is rapidly evolving with fact checks playing a new important role. However, the ecosystem is still missing an important piece of functionality --- not surprisingly, articles with fake or controversial news do not link to fact checks that discuss their veracity. Since fake news mutates and spreads quickly to different sites, fact checking sites can't link to all the pages carrying the dubious claims. This missing link reduces the potential impact of fact checks. Without an easy way of going from a news article to a fact check that investigates it, fake news gains traction.


In this paper, we propose a new `Related Fact Checks' (henceforth referred to as RFC) service. Consider the following scenario: a user is  reading an article which makes claims that they would like to verify. Currently, they would have to go to Bing, Google or another search engine and do a set of searches constructed out of the terms/entities in the article, hoping to find a relevant fact check. Imagine instead, a service that would retrieve relevant fact checks, if any. This service would not decide if the claims were true or false. Rather, it would provide a short list of relevant fact checks available and let the user make an informed decision.  Further, the service could provide links to other articles, that are about the same topic from the same site, that have been fact checked. The RFC service, which could be made easily accessible through a browser extension, would do the equivalent of adding a hyperlink from a fake news article to fact checks that investigated it.

 Given an article, the goal of the RFC service is to find the fact check(s) that address the claims made by the article. This task is complicated by the following phenomenon: It is far cheaper and easier to concoct fake stories than it is to fact check them. Consequently, the vast majority of fake stories  have not been fact checked. However, we note that fake stories follow certain patterns. In particular, our analysis (see figure \ref{fakehistogram}) shows that there are a small number of themes (e.g., anti-vaccine, anti-climate, etc.) that appear repeatedly in many fake stories. So, even if the article the user is reading does not have a corresponding fact check, showing the user fact checks for stories that are on the same theme will help the user more critically interpret the article. 

From an algorithmic perspective, this distinguishes the RFC service from traditional information retrieval. In addition to knowing term distribution statistics across the corpus, we need to discover the themes that occur in the corpus and exploit this structure for the retrieval and ranking. 




The main contributions of this paper are as follows. We introduce the concept of `Related Fact Checks' service as a means of combating fake news. We discuss the core information retrieval problems such a service needs to solve and propose techniques for these. In particular, we discuss how we may identify long-running themes and exploit them in retrieving relevant fact checks.  We discuss our implementation of this service, describe the browser extension through which the service is accessed, and present precision and recall numbers from an evaluation of the results provided by the service.



\subsection{Related Work}


 The topic of computational treatments of fake news has received a substantial interest in the research community recently. Conroy, Rubin, et al. \cite{conroy2015automatic} survey the  landscape of veracity (or deception) assessment methods, their major classes and goals, all with the aim of proposing a hybrid approach to system design. As they describe, there are two major categories of methods: (1) Linguistic Approaches in which the content of deceptive messages is extracted and analyzed to associate language patterns with deception; (2) Network Approaches in which network information, such as message meta-data or structured knowledge network queries can be harnessed to provide aggregate deception measures. Chiu, Gokcen et al. \cite{chiu2013classification} examine different ways of classifying fake and real articles using Support Vector Machines. 

 Much of the work on the spread of fake news has focused on its dissemination through social media. Shao, et al. \cite{socialbots} show how bots on social networks have contributed greatly to the spread of fake news.
Jin and Dougherty \cite{Jin:2013:EMN:2501025.2501027} apply epidemiological models to information diffusion on Twitter. This paper is the first to employ the SIEZ model to Twitter data and shows the success of this method in capturing the spread of information on Twitter. Tacchini and Ballarin \cite{DBLP:journals/corr/TacchiniBVMA17} shows that Facebook posts can be determined to be hoaxes or real based on the number of likes.  The authors use two classification techniques; one is based on logistic regression while the other on a novel adaptation of boolean crowdsourcing algorithms. Gupta, Lamba, et al. \cite{Gupta:2013:FSC:2487788.2488033} show how a small number of users were responsible for a large number of retweets of fake images of Hurricane Sandy. Gupta, et al. \cite{Gupta:2012:CRT:2185354.2185356} used regression analysis to identify the important features which predict credibility. The authors used machine learning to create an algorithm which ranked tweets based on the credibility of sources. 

The past year has seen numerous announcements of `Fake News Challenges', in which programs compete to automatically identify fake news \cite{fakenewschallenge}, \cite{kagglefakenews}. Our approach, in contrast, does not attempt to tell the user whether an article is true, or even to what degree it is true. Instead, we aim to supplement the article the user is reading with fact checks which might enable the user to more critically interpret the article. 
To our knowledge, this is the first work on providing a service that automatically provides connections between articles that might potentially be fake news and fact checks.

Most previous work in Information Retrieval is based on features corresponding to the frequency of occurrence of terms /n-grams in a document and in the corpus. Kurland and Lee \cite{kurland} introduced the idea of using the structure of the corpus to help with ad-hoc queries. In their work, they used clustering to identify the structure in the corpus. Long-running themes are one kind of structure that may occur in corpora of news articles. Here, we use Topic Modeling for identifying
these themes, and then use these themes to retrieve relevant fact checks. We believe that this approach has wide applicability in Information Retrieval, beyond fake news and fact checks.

Recent studies in the field of Political Science  suggest that exposing readers to fact checks has a substantial positive impact in the long run. Hill \cite{doi:10.1086/692739} finds empirical evidence that voters do gradually change their opinions when presented with the facts. Similarly, Peterson \cite{doi:10.1086/692740} finds that voters with more information are less likely to vote along party lines. Pennycook et al. \cite{pennycook2017falls} investigate characteristics of readers who believe in fake news and find an inverse correlation between critical thinking abilities and the likelihood of believing in fake news. Finally, Brandtzaeg et al. \cite{brandtzaeg2017trust} study the perceived trustworthiness of fact check services themselves and conclude that fact checking services need to increase their transparency by disclosing their methodologies and funding sources.

\section{Methodology}

 We now discuss the problems involved in creating a Related Fact Check service:
 \begin{enumerate}
 \item Create a corpus of fact check articles
 \item Identify long-running themes in fake news stories
 \item Rank fact checks based on their relevance to a given article
 \item Map from a given article to other articles on a particular site, that are similar and have been fact checked
 \item Create a user interface for accessing the service
 \item Evaluate the relevance of results
 \end{enumerate}

\subsection{Creating a fact check corpus}
\begin{figure*}
\begin{center}
  \includegraphics[scale=.6]{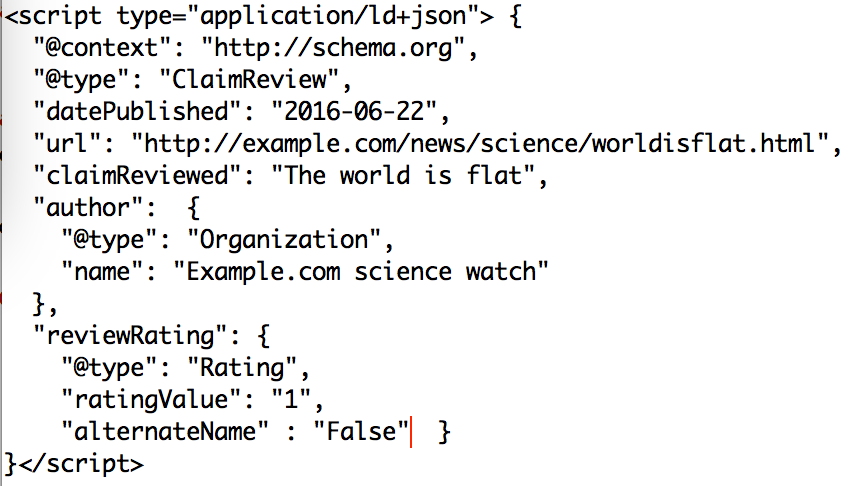}
  \caption{Example of ClaimReview markup in Schema.org vocabulary (source \cite{claimReviewMarkup})}
  \label{schema_markup}
  \end{center}
\end{figure*}

In order to retrieve the fact checks relevant to a given article, we need a corpus of fact checks. According to the Duke Reporter's Lab, there are about 116 sources providing fact checks today \cite{dukereporterslab}. In order to determine the relevance of a fact check to a given article, we need to identify the primary claim being investigated. Extracting the claim from the body of the fact check content can be difficult and imprecise. Instead, we rely on the Schema.org ClaimReview  markup which gives us the reviewed claim. A majority of the sites that provide fact checks now carry this markup on their pages. See Table \ref{claimReviewed} for some sample values of the claimReviewed field. Fig. \ref{schema_markup} gives an example of such a markup that may be found on a fact check page.

For the version of the service described in this paper, we restricted our attention to fact checks in English from sites that provide the Schema.org markup. We crawled the pages on these websites and extracted the ClaimReview data. After removing duplicates, we built a corpus of 5350 fact checks from 45 sites. For each fact check, in addition to the URL, the markup gave us the title of the fact check, the specific claim that was reviewed, the date of the fact, and how the claim was rated by the article. 
 
\begin{table}
\begin{center}
\begin{tabular}{|l|}
\hline
A damaged nuclear reactor at Fukushima \\ Daiichi is 
about to fall into the ocean. \\
\hline
Australia is the first country to begin \\ 
microchipping its citizens. \\
\hline 
There are "no-go zones" in Sweden where the \\ 
police can't enter. \\
\hline
A federal judge ruling in a defamation suit \\
declared  that CNN was "fake news." \\
\hline
\end{tabular}
\end{center}
\caption{Sample of some of the claims reviewed in the fact checks. Source \cite{politifact}, \cite{washingtonpost}}
\label{claimReviewed}
\end{table}
 




\subsection{Extracting Themes in Fake News}

\begin{figure*}
\begin{center}
  \includegraphics[scale=.6]{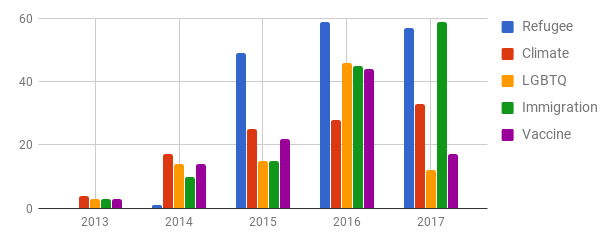}
  \caption{Distribution of fake news articles for 5 themes, over 5 years.}
  \label{fakehistogram}
  \end{center}
\end{figure*}

When we examine a corpus of fake news articles, we find that there are certain long-running {\it themes} with a number of stories around each theme. Each story within a specific theme may have a different claim but will still relate to the theme. 
For example, one prominent theme amongst fake news articles is that vaccines are harmful. There are many specific stories that fall into the genre of anti-vaccine stories, including: `CDC raided by FBI to seize data on vaccines' \cite{f1}, `If a vial of vaccine is broken, the building must be evacuated' \cite{f2}, `HPV vaccine causes death of 32 year old' \cite{f3}, etc. Our analysis indicates that there are at least 80 stories related to vaccines. Another recurring theme in fake news is about climate change. There is a wide range of stories related to this theme, ranging from those about the climate, such as 'Carbon dioxide is not a primary contributor to the global warming that we see'\cite{f5} to those that are related to the political side of climate change, such as `California legislators have made it illegal for anyone to deny climate change, under threat of jail time' \cite{f4}. To validate our hypothesis of long-running themes in fake news, we identified five such themes and built classifiers for each theme. We collected a corpus (using a procedure described later) of articles from sites that often publish fake news and ran these classifiers on this corpus. Fig. \ref{fakehistogram} shows the distribution of these 5 themes over 5 years. 

\subsection{Themes and Relevance} 
When there is a fact check that is very specific to the claim being made in the article, clearly, RFC should show the user that fact check. However, even when there is no fact check that addresses the specific claim made by the article, if the article falls into a theme where other stories have been fact checked, it would be useful to provide the user with those fact checks. These related fact checks will hopefully enable the user to more critically interpret the article. For example, there is a story published by the site `Daily Sheeple' with the headlines `BOMBSHELL: CDC Commits New Vaccine-Autism Crime - Won't Allow Whistleblower to Testify' \cite{f6}. As of September 2017, this specific story does not have a fact check. However, stories that are related to this general theme, such as the one which claimed that the FBI raided the CDC to seize data about vaccines, do have fact checks. If the user can be shown that this new story is just the next iteration in a long series of stories, many of which have been shown to be not true, the user will be in a better position to judge the veracity of the article they are reading.

We recognize two distinct levels of relevance of an article to a fact check. At the first level are those which specifically address the claims made by the article. At the second level of relevance are the fact checks that don't address the specific claim, but addresses claims that are about the same theme and related to the claim made by the article. We are interested in identifying both kinds of fact checks.  

\subsection{Automatically identifying themes}
 We manually identified the themes and created classifiers for the themes in figure \ref{fakehistogram}. This is clearly not scalable when the number of themes is much larger and the corpus is evolving. We would like a scalable way of identifying themes. We now discuss two approaches for doing this: clustering and topic modeling.

\subsection{Clustering vs. Topic Modeling}
 Clustering is one of the most commonly used techniques for identifying patterns in data sets. Kurland and Lee \cite{kurland} use clustering in their work on using corpus structure for search. We experimented with clustering to identify themes, using K-means implementation in the Scikit package \cite{scikit-learn}. We discovered a problem with relying on clustering, which stems from the fact that each article belongs to exactly one cluster. Given an article such as \cite{syriavaccine} that covers the vaccination related deaths of several kids in Syria, with clustering, this can either belong to a vaccine cluster or Syria cluster. We need a model that associates the article with both themes. 
 
Starting with Blei, Jordan, et al. \cite{blei2003latent} there has been substantial work on `Topic Modeling', which identifies a set of `Topics' which can be combined in different proportions to generate the articles in a corpus. Topic Modeling has become an effective tool for the discovery of underlying semantic themes in document corpora.  In Topic Modeling, each topic is a set of words, with a weight associated with each word and each article is modeled as being composed of a set of topics in different proportions. This makes it very natural for an article to be part of multiple themes.

\subsection{Capturing themes with Topic Modeling}
In this work, we use Topic Modeling as a tool for identifying recurring themes in fake news stories. Our goal of identifying themes in fake news is complicated by the fact that we do not have a comprehensive corpus of fake news stories. We note that the fact checks themselves reflect these recurring themes. For example, we find a number of fact checks corresponding to the theme of vaccines being harmful, another set corresponding to the theme of climate change being a hoax and so on. We identify the major themes in fake news by applying Topic Modeling on the corpus of fact checks.

We use the ldaModel class in the Gensim \cite{gensim} tool. We preprocess our corpus to remove all words that appeared in more than half the documents (which eliminates stop words, etc.). With Topic Modeling, one of the input parameters is the number of topics desired, which is a function of the corpus. In our case, after experimentation, we settled on 300 topics. Once we compute our model, we can use it to assign topics to any document --- article or fact check. Given a bag of words from the document --- words from the title, the claim reviewed or the content of the article --- the model can be run to give us a set of topics (and relative proportions) corresponding to that bag of words. The number of topics (and their proportions) shared by an article and fact check can be viewed as a measure of their similarity.

\subsection{Identifying Topics corresponding to themes} 
Each article is modeled as being composed of a set of topics, in different proportions. Each topic can be thought of as capturing some aspect of the document. By removing the most commonly occurring words, we eliminated the topics that capture the basic elements of language, such as pronouns, articles, common words, etc. Some  topics  capture stylistic aspects of particular publishers. For example, a fact check from Washington Post has certain stylistic elements that are not found in a fact check from a publication such as Snopes. Such topics, which capture such stylistic elements and which are part of most documents, don't reflect themes in fake news. We do find that some number of topics correspond to the kind of themes we are aiming to capture. Clearly, an article and fact check matching on one of these topics that correspond to a theme is more significant than their matching on one of the topics corresponding to stylistic elements.


 We manually went through the top five documents associated with each topic and identified 18 topics that clearly correspond to themes. Table \ref{thematictopics} lists some of these topics. Table \ref{topictable} lists some topics that were not identified as being thematic. It should be noted that the actual terms in these topics are stemmed, but for the sake of readability, we have used one of the unstemmed versions of the words in these tables. 
\begin{table}
\begin{center}
\begin{tabular}{| l | l |}

\hline
Name & Top Words \\

\hline
Vaccines & vaccine, infant, flu, cdc, monument, \\
& hpv, gardasil, pediatrics, autism, ...\\
       
\hline
Climate & climate, carbon, emission, pollute,\\
& epa, co2, fossil, earth, atmosphere,  ...\\

\hline
LGBTQ    & gender, bathroom, transgender, \\
& gay, discriminate, lgbt, ... \\


\hline
\end{tabular}
\caption{The top words associated with a few of the topics corresponding to fake news themes. The names were manually assigned. }
\label{thematictopics}
\end{center}
\end{table}

\begin{table}
\begin{center}
\begin{tabular}{|l | l |}
\hline
Topic Number & Top Words \\
\hline
Topic 32 & deadbeat, pure, clue, five, exist, \\
& none, pig, insert, hud, poppi, ...\\
      
\hline    
Topic 44 & none, peanut, morgan, lynch, \\
& hr, ingredient, trace,  \\ 
        
\hline          
Topic 45 & none, mail, satan, injury, ration, \\
& singer, sir, temple, microphone, \\ 
         
\hline
\end{tabular}

\caption{Sample of topics that don't seem to correspond to themes in the fact checks.}
\label{topictable}
\end{center}
\end{table}

\subsection{Related Articles Corpus}

 As shown in figure \ref{screenshot2}, when the user retrieves fact checks related to a particular article, we also show a few other stories that are similar, from the same site as the article, which have a high relevance to one of the fact checks being displayed (i.e., appear to have been fact checked). If there is a cluster of articles on that site, that is along the same theme, at least some of which have been investigated, the user can get a better idea of the overall veracity of that site, with respect to this topic. 
 
 The problem of generating the related articles  is the inverse of the RFC service. With RFC, we go from an article to fact checks. In contrast, here we are given a fact check and want to find stories that carry the claim discussed by the fact check. Since the total number of fact checks is fairly small, we compute articles related to a fact check in advance. We now discuss how we compute this corpus of related articles.

As discussed earlier, with each fact check we have a short statement of the claim that is reviewed by the fact check (see Table \ref{claimReviewed} for example). We construct shorter queries from the claimReview strings (by removing stop words, extracting named entities, etc.) and issued these as queries to a web search engine (Google), collect the first 20 results from each query and collate these result by the site. Because of the nature of web search, most of the result  pages do not  make the claim stated in the claimReview, but are pages from well-known sites (such as Wikipedia) that mention some of the terms in the claimReview.
We examined a random sample of pages from each of the sites in the set of results and manually identify those sites that repeatedly touch on stories that have been fact checked. These are the sites that tend to carry articles that often make dubious claims. We created a Google Custom Search Engine which restricts results to pages from these sites. This search engine is optimized for retrieving articles likely to be fact checked. We issued the same queries (derived from the claimReview statements) to this Google Custom Search Engine. We crawled a set of 5,000 result pages. This gives us a set of articles from these sites for each fact check. This is our related stories corpus. We also use this corpus for tuning the weights in the scoring function and for evaluating the quality of the results.

\subsection{Accessing the RFC service}
\begin{figure*}[!htb]
\begin{center}
  \includegraphics[scale=.6]{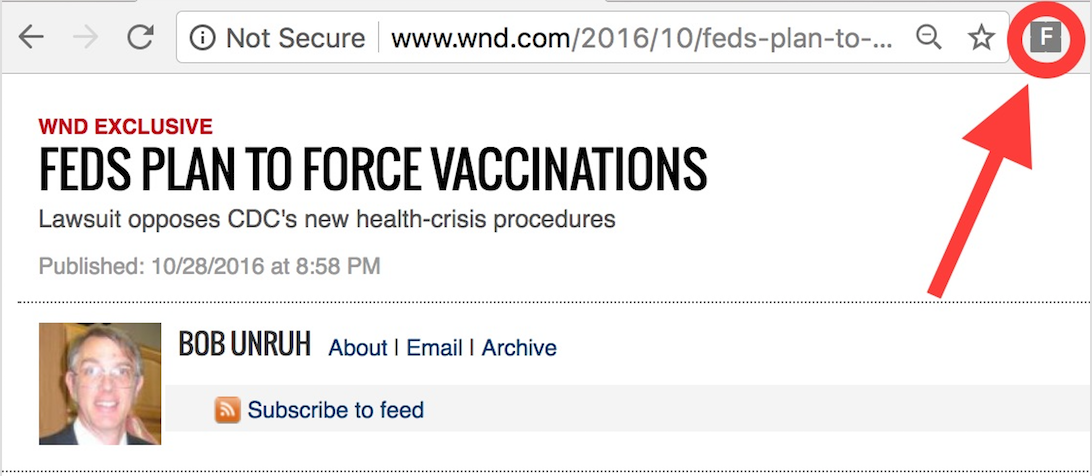}
  \caption{Browser with extension, showing icon for accessing the extension, through which the Related Fact Check service is invoked. }
  \label{screenshot1}
  \end{center}
\end{figure*}

\begin{figure*}[!htb]
\begin{center}
  \includegraphics[scale=.6]{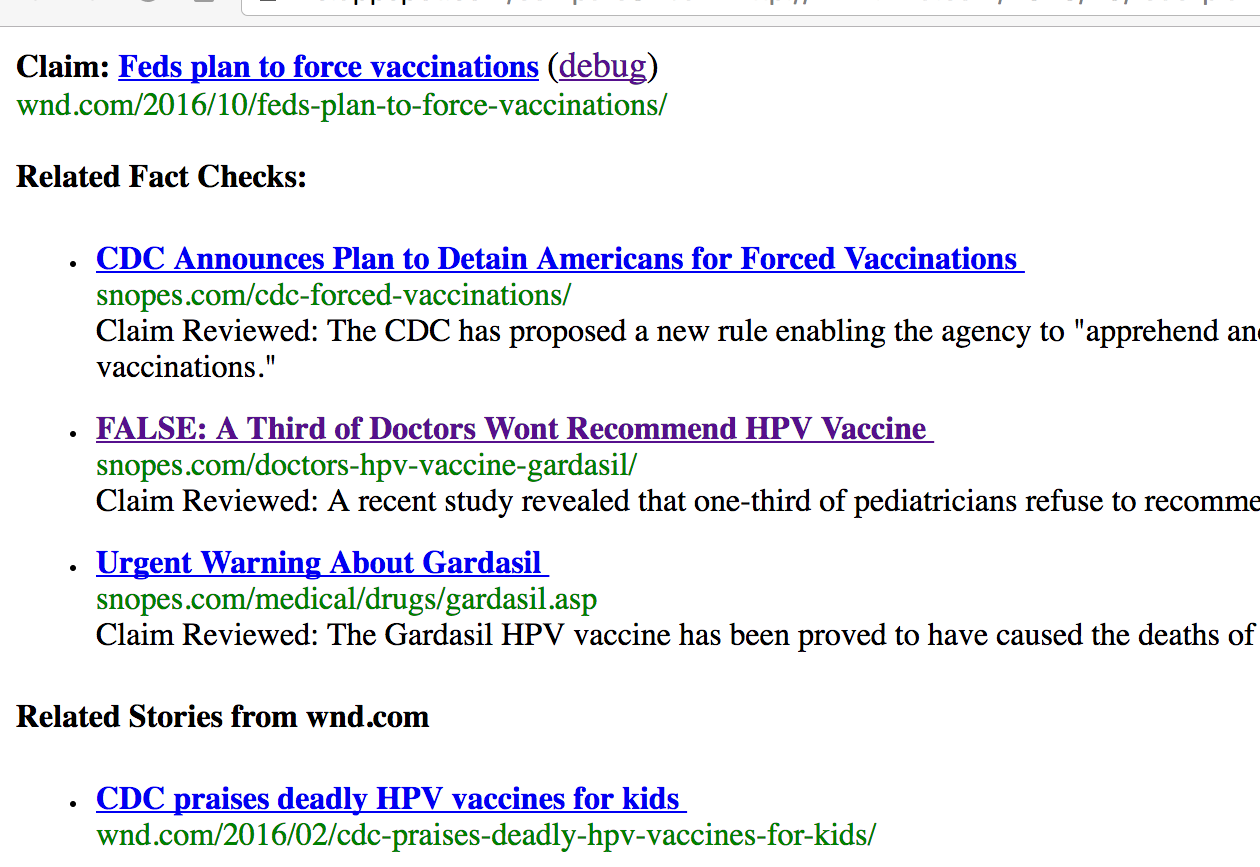}
  \caption{Showing the presentation of the Related Fact Checks and Related stories for article in figure \ref{screenshot1}, using the extension and Related Fact Check Service.}
  \label{screenshot2}
 \end{center}
\end{figure*}

 The service can be accessed with a browser extension. When the user is on a page, they can request fact checks related to that article by clicking on the `F' icon next to the location bar in her browser. The results are presented in a very simple interface. The other stories related to one of the fact checks retrieved from the same site as the original article, if any, are listed below the fact checks.  Figures \ref{screenshot1} and \ref{screenshot2} show screenshots of the extension and the page of fact checks and related articles shown to the user when they click on the `F' icon.
 
\subsection{Ranking the relevance of Factchecks}

Given an article, we would like to find relevant fact checks if they exist. This problem bears many similarities with traditional Information Retrieval (text search) and many of the techniques developed in that field can be applied here. We use the Vector Space model \cite{salton}, in which each document is modeled as a vector in a space where each term corresponds to a dimension. The coordinate of a document along a term/dimension is the frequency of that term in the document (TF), scaled by the number of documents the term appears in (IDF). Similarity between two documents is measured by the cosine of the angle between the vectors corresponding to two documents. There are some important distinctions between traditional search and our application that guide our adaptation of the vector space approach. Specifically,
\begin{enumerate}
\item Our primary goal is to find the fact that addresses the claim made by the article. The claim checked by a fact check is available  via the ClaimReview markup. While we don't have a similar markup for the article, the title often captures an article's main claim. Therefore, similarity between the title of the article and claim review of the fact check should be given special significance.
\item As discussed earlier, we compute the topics corresponding to the article and to each fact check. The set of possible topics can be modeled as additional dimensions in the vector space and similarity along these dimensions, especially along dimensions corresponding to topics that map to themes should be given their own weight in the ranking.
\item In typical search, the goal is to retrieve some number of the most relevant documents, even if some of them are not particularly relevant to the query. I.e., typically, there is not a hard cutoff for the relevance, where we don't show documents that fall below a certain threshold. In the context of a fact check retrieval service, we need such a cutoff. Showing the user irrelevant fact checks is likely to make the user less likely to use the service and should hence be avoided. So, we also have to compute a threshold so that fact checks which have a score below that threshold are not returned.
\end{enumerate}
The following formula is used to compute a score measuring the relevance of an article $a_i$ to a fact check $fc_j$
\begin{eqnarray}
S_{a_i, fc_j} = &  W_{title} * S_{title} \\
			    &+ \hspace{.1in} W_{body} * S_{body} + \nonumber\\
                & +W_{topics} * S_{topics} &   \nonumber \\
                 &+W_{thematic-topics} * W_{thematic-topics}  &\nonumber
\end{eqnarray}

where $S_{title}$ is the cosine of the angle between TFIDF weighted vectors corresponding to the title/claim review of the article/fact check, $S_{body}$ is the cosine of the angle between TFIDF weighted vectors corresponding to the body of the article/fact check, $S_{topics}$ is the cosine of the angle between the vectors in topic space between the article and factcheck, weighted by the topics proportion and $S_{thematic-topics}$ is the  corresponding to the cosine of the angle between the vectors in thematic topic space between the article and factcheck, weighted by the topics proportion. $W_{title}$, $W_{body}$, $W_{topics}$ and $W_{thematic-topics}$ are the weights associated with these scores. We also employ a cutoff $T_l$ such that when the score is below this threshold, we don't display the corresponding fact check. 

\subsection{Relevance and tuning}

For a given article, a fact check can be one of the following:

\begin{itemize}
 \item On Claim: The fact check addresses (one of) the main claim(s) of the article. 
 \item On Theme: The fact check does not address the main story of the article, but addresses other stories that are in the same storyline or theme of the article. These fact checks would help the reader understand the article and place it in context.
 \item Irrelevant: The fact check is irrelevant to the article. Presenting the reader with this fact check is likely to make them not ask for relevant fact checks in the future.
 \end{itemize}

 For example, given an article published by the Observer claiming `The Clinton Foundation Shuts Down Clinton Global Initiative' \cite{f6}, the fact check from FactCheck.org checking the claim `The Clinton Foundation is shutting down due to lack of donations' \cite{f7} would be considered `On Claim'. A fact check from PolitiFact about the salaries of the Clintons for the Clinton Foundation  \cite{f8} would be considered `On Theme' as it could help the user place the article in context. Lastly, fact checks that have no association with the topic such as those concerning health care would be considered `Irrelevant'.
 
 The goal is to retrieve all the `On Claim' fact checks, some of the `On Theme' fact checks and none of the `Irrelevant' fact checks. In traditional information retrieval, where there are only two categories of results --- relevant and irrelevant, typically, the system is optimized for the F-measure, the harmonic mean between the precision and recall. Given that we have two categories of relevant results, we adapt that as follows. We assign a score of 2 for every `On Claim' result, a score of 1 for every `On Theme' result and a score of -2 for every `Irrelevant Result'. The cumulative score for a set of results is simply the sum of the scores for each result. Our goal is to pick a set of values for the weights in the earlier equation to maximize this cumulative score over a set of articles. We picked 20 articles from the related stories' corpus, generated the top five related fact checks for different values of the weights, evaluated each result 
(i.e., assigned it a score of 2,1 or -2). We picked a set of weights that 
maximized the net cumulative score for these 20 articles.

\section{Results \& Discussion}
 We randomly picked 100 articles from the related stories corpus and evaluated the performance of the different features and their combination. For each article, we computed five of the most related fact checks whose score was above a certain threshold. In some cases, there were less than five fact checks above the threshold. We manually evaluated the results and determined if  each article and fact check pair was `On Claim', `On Theme' or `Irrelevant'.
 
 Most pages on the web don't have a fact check associated with them. However, because of the way we constructed our related stories corpus, most, but not all, of the pages have either an `On Claim' fact check or `On Theme, fact check. In an ideal case, we should retrieve at least one `On Claim' fact check, if there are any, and one or more `On Theme' fact checks, if there are any, and no `Irrelevant' fact checks. Results of the evaluation are presented in the next.
 
\subsection{Evaluation Results} 
Table \ref{results1} gives the precision of the results from the evaluation of the 100 randomly chosen stories. In addition to the scoring formula given earlier, we present the results obtained from each of the features independently to better understand the contribution of each one. Note that the total number of results for different feature combinations vary. 
That is because only results whose score is above a certain threshold are displayed. Each of the features and the combination of features has its own threshold. 
Table \ref{results2} is a measure of recall and gives us the fraction of pages for which there is an "On Claim" result which was retrieved. 


\begin{table}
\begin{center}
\begin{tabular}{| l | l | l | l | }
\hline
Features & On Claim & On Theme & Irrelevant \\
\hline
Title / Claim review & 0.24 (91) & 0.23 (99)  & 0.52 (202)  \\
\hline
Page Content & 0.25 (128) & 0.30 (188)  & 0.43 (219) \\
\hline
Topics& 0.10 (67) & 0.25 (153) & 0.63 (310) \\
\hline
Thematic topics   & 0.11 (65) & 0.35 (182) & 0.53 (257) \\
\hline
 All features & 0.23 (130)  & 0.41 (232) & 0.35 (198)  \\
\hline
\end{tabular}
\caption{Fraction i.e., precision and absolute number (in parenthesis) of results  in each category for each set of features. Note that the total number of results for different features are different because of differing thresholds.}
\label{results1}
\end{center}
\end{table}

\begin{table}
\begin{center}
\begin{tabular}{|l | c|}
\hline
Features & On Claim Recall \\
\hline
Title / Claim review& 0.72 \\
\hline 
Page Content & 0.85 \\
\hline
Topic match & 0.62 \\
\hline
Thematic topic match   & 0.55  \\
\hline
All features & 0.88  \\
\hline
\end{tabular}
\caption{Fraction of articles (i.e., recall) that have `On Claim' fact checks for which at least one of these fact checks was retrieved.  }
\label{results2}
\end{center}
\end{table}

From table \ref{results1}, we see that when we use all our features, we are able to retrieve both `On Claim' and `On Theme' results 
a very high fraction of the time. We can also see that the fraction of pages for which at least one `On Claim' result is retrieved, when there is one, is quite high. However, we note that the number of `Irrelevant' results is still quite high.
 
While content-based matching can be effective, `On Theme' improves with the addition of topics. Surprisingly, just using topics, with and without extra weighting for topics corresponding to themes, performs as well as just the content of article for retrieving  `On Topic' as content, but not on "On Claim". Just using topics doesn't do as well on `On Claim', because though topics tend to be a good distillation of the overall content, they do poorly on capturing the precise semantics of a particular claim. Thematic topics give fewer off-topic results, since some of the topics tend to capture aspects of documents that have more to do with the writing style of different publications than with the content of the story.

  Matching only on titles (and claimReviewed) does better than expected, likely because a number of titles  provide a good synopsis of the article. However we see a category of poor retrieval that arises out of the fact that many fake articles tend to  use titles that are targeted more at attracting clicks than conveying a summary of the article, e.g., articles with titles such as "You won't believe who just endorsed ...", where the title similarity is less effective than the content similarity.
  
\section{Conclusions and Future Work}
 In this paper, we introduced a new service, RFC, for helping users make better judgments about articles they read, especially when they make claims that might seem fake. Given the scale and scope of fake news, it is clear that we need computational tools to combat this phenomenon. The problem is more nuanced than simply training classifiers that classify articles as being fake or true because there are many stories that occupy a gray zone. The goal should be to give the user context so that they can critically read an article. The emergence of  fact checking organizations and their extensive adoption of Schema.org schemas offers an opportunity for a new kind of service. To our knowledge, this is the first service of this kind. 
 
  We introduce the concept of a `Related Facts Checks' service that enables a user to get a set of fact checks that may be relevant to an article that they are reading. We describe how such a service may be built and present results from an early implementation of this service. As demonstrated by the evaluations, even our early implementation offers results that could be helpful to a user.
  
  The work presented here is just a first step and opens up many new directions. In particular, our implementation uses a number of parameters/weights for combining different features. In the initial implementation discussed in this paper, these weights were manually adjusted. In the future,  we hope to explore the use of machine learning to automatically tune the system.
  
  The approach described in this paper made significant use of Topic Modeling. Today's Topic Modeling tools generate many topics for a given corpus. We showed how identifying and giving greater weight to thematically more meaningful topics improves the performance of the system. In this work, we identified such topics manually, which poses a challenge as the number and scope of fact checks increases. Another direction of future work is to automate the identification of the topics that satisfy this condition.
  
  It would be useful to indicate to the user that the article being read may have related fact checks. To do this, the service needs to check for related fact checks even before the user clicks on the browser extension icon. However, contacting a remote server every time the user visits any page on the web is not only computationally expensive, but also a potential privacy violation. As discussed earlier, matching just on titles/claimReviewed yields surprisingly good results. And since titles/claimReviewed strings are small, they can be bundled into the browser extension, enabling a lightweight service that can inform the user that related fact checks may be available by changing the color of the browser extension icon. And since all the computation takes place in the browser, this can be done without loss of privacy.  Further, title matching can be improved by the use of word similarity metrics. For example, with strict matching, the words `immigrant' and `refugee' will not match. However, the two words are more similar to each other (and almost replaceable in the language of article titles) than most other words. We are currently investigating the use of embedding techniques (such as Word2Vec) for doing this.

 We spent a significant amount of effort curating the corpus of fact checks and articles. We believe that the availability of open dataset, preferably one that allows others to contribute to it, will greatly facilitate research on this important topic. In that spirit, we plan to make the data collected publicly available.

\bibliography{sample}
\bibliographystyle{aaai}

\end{document}